# Analysis of social interactions in group-housed animals using dyadic linear models

Junjie Han[1,2], Janice Siegford[1], Gustavo de los Campos[3,4,5], Robert J. Tempelman[1], Cedric Gondro[1], and Juan P. Steibel[1,6,*]

[1]Department of Animal Science, Michigan State University, East Lansing, MI 48824, USA;

[2]Department of Computational Mathematics, Science and Engineering, Michigan State University, East Lansing, MI 48824, USA;

[3]Department of Epidemiology and Biostatistics, Michigan State University, East Lansing, MI 48824, USA;

[4]Department of Statistics and Probability, Michigan State University, East Lansing, MI 48824, USA;

[5]Institute for Quantitative Health Science and Engineering, Michigan State University, East Lansing, MI 48824, USA;

[6]Department of Fisheries and Wildlife, Michigan State University, East Lansing, MI 48824, USA

[*]Corresponding author. E-mail address: steibelj@msu.edu (Juan P. Steibel).

**Abstract**

Understanding factors affecting social interactions among animals is important for applied animal behavior research. Thus, there is a need to elicit statistical models to analyze data collected from pairwise behavioral interactions. In this study, we propose treating social interaction data as dyadic observations and propose a statistical model for their analysis. We performed posterior predictive checks of the model through different validation strategies: stratified 5-fold random cross-validation, block-by-social-group cross-validation, and block-by-focal-animals validation. The proposed model was applied to a pig behavior dataset collected from 797 growing pigs freshly remixed into 59 social groups that resulted in 10,032 records of directional dyadic interactions. The response variable was the duration in seconds that each animal spent delivering attacks on another group mate. Generalized linear mixed models were fitted. Fixed effects included sex, individual weight, prior nursery mate experience, and prior littermate experience of the two pigs in the dyad. Random effects included aggression giver, aggression receiver, dyad, and social group. A Bayesian framework was utilized for parameter estimation and posterior predictive model checking. Prior nursery mate experience was the only significant fixed effect. In addition, a weak but significant correlation between the random giver effect and the random receiver effect was obtained when analyzing the attacking duration. The predictive performance of the model varied depending on the validation strategy, with substantially lower performance from the block-by-social-group strategy than other validation strategies. Collectively, this paper demonstrates a statistical model to analyze interactive animal behaviors, particularly dyadic interactions.

# 1. Introduction

The study of social interactions is of paramount importance in applied animal behavior research (Rodenburg et al., 2010; Silk et al., 2018). Researchers are interested in elucidating the basis for the observed variation in the intensity and frequency of interactions among pairs of individuals that are part of a social group. Some of the applications of such study include mate choice (Andersson and Simmons, 2006; Bierbach et al., 2013), aggression and other damaging behaviors (Angarita et al., 2019; Oczak et al., 2013; Peden et al., 2018), and competition for access to feeding space (Angarita et al., 2021; Lu et al., 2017), etc. Thus, given data on pairwise behavioral interactions recorded from an experimental or observational study, it is necessary to quantify the effects of various individual- and group-level factors on social interactions.

Data from pairwise social interactions are considered dyadic (Kenny et al., 2020). This is, the unit of observation is not the individual, but a pair of individuals. In general, dyadic interaction data can be arranged in square matrices. It can be further re-arranged in the form of a response vector, which is generally accomplished in two different ways (Figure 1): a) The data are summed row-wise/column-wise to represent an individual level-observation (total duration that each animal is engaged in a particular behavior regardless of whom the animal interacted with), or b) the matrix elements are stacked keeping intact their dyadic nature. In the first case, there is loss of information, and it should be avoided if the aim is to study the dyadic nature of social interactions. In the second case, however, it is of utmost importance that all sources of variations are modeled to properly account for group means, variances and covariances between subsets of the data (Figure 2); otherwise, if important factors are ignored, this can adversely affect estimates and predictions.

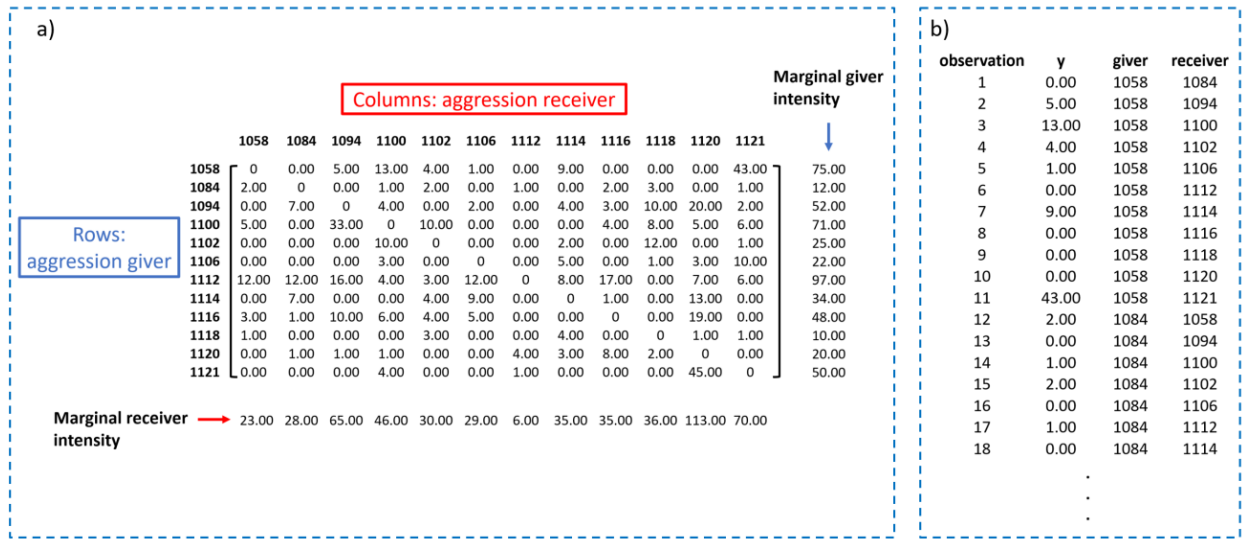

**Figure 1** Panel a), directional dyadic interaction intensity matrix (elements in the matrix represent attacking duration); row sums and column sums are shown in the margins of the matrix. Panel b), a truncated long-format table that is re-arranged from the interaction matrix; each row represents a record that is the attacking duration in seconds from a giver animal to a receiver animal. 0.00 means observed zero while 0 means structural zero that we do not consider as an actual interaction.

A proper way to model dyadic data is to fit generalized linear mixed models (GLMM) that include fixed and random effects to account for means and covariances depending on the actual design of the experiment (Kenny et al., 2020). In this study we describe how GLMM can be used to analyze dyadic data and illustrate how to use this approach to analyze a pig behavior dataset. First, we defined the type of social interaction data and how to properly model variation in the response using GLMM. Second, we applied the proposed GLMM to the experimental data and illustrated how to elicit, fit, and check the models and how to interpret the results. Finally, we performed posterior predictive checks of the models through several validation strategies. The GLMM presented in this paper can be used by applied animal behaviorists to analyze other pairwise social interaction data to obtain statistically valid and biologically meaningful results, which can be helpful to understand interactive behaviors of animals for practical purposes of management or improved welfare.

## 2. Methods and Materials

### 2.1 Data from social interactions should be analyzed as dyadic data

For a social interaction to occur, at least two animals need to be involved. Although behavioral interactions may involve more than two animals at a time, in this paper we assume that the data on social interactions is obtained through observations of pairwise/dyadic behaviors and that it can be arranged in an interaction matrix (Figure 1a).

The data may be obtained within a single large social group, in which all the potential pairwise interactions have been monitored and quantified. Alternatively, the dyadic data can be collected from several social groups of variable sizes, within which all potential pairwise interactions have been monitored and quantified, but no between-group relations are possible.

We also assume that in addition to the social interactions per se other variables have been observed. These variables may be individual-specific or dyad-specific. Examples of individual-specific variables are those related to each individual's age, sex, size, and past life experiences (e.g., early-life social or nutritional stress) and they can be continuous, discrete or categorical in nature. Dyad-level variables are those that only pertain to the pair of individuals. For instance, a dyad-level variable can be described as whether they have met each other before the interaction is observed. It is important to notice that sometimes individual-level variables may be coded as dyad specific, for instance, the difference in live weight between two animals can be viewed as a dyadic-level observation, but in fact it arises from a linear combination of two individual-level variables. In that case, we prefer to keep individual level observations separate.

#### 2.1.1 Social interaction data

Social interaction data can be of different types. From a mathematical point of view, the social interactions could be represented by a binary outcome (0/1=it occurred/it did not occur), by a discrete outcome (frequency of occurrence of an interaction), by an ordinal outcome (intensity or severity of interaction on an arbitrary scale), or by a continuous outcome (intensity of the interaction on a continuous scale, duration of the interaction, etc.). The practical implications of the different types of responses pertain to the statistical distribution that is used to model the stochasticity in social interaction data.

From the point of view of the directionality of the behavior, in most cases, we can assume that the behavior is directional i.e., there is a giver and a receiver. For instance, in the study of animal aggression, in many cases there is a clear attacker and a victim. In studies of feather pecking in group-caged chickens (Savory and Mann, 1997) and tail-biting in group-housed pigs (Angarita et al., 2019; Wurtz et al., 2017), there was one animal that was delivering the behavior (we will call this animal the giver animal) and another one which was clearly receiving the behavior (the receiver). In the following subsections we lay out these concepts with the directional interaction data, and we summarize the model parameterization in a generalized form.

### 2.1.2 Analysis of dyadic data from directional social interactions

When the social interactions are directional, the data collected from each social group can be arranged in a matrix as represented in Figure 1a. If there are $n$ animals in a certain group, $n(n-1)$ interactions will be observed within the group. We assume that there is no measurement error implying that if an interaction was recorded then this interaction did indeed happen as recorded, and, perhaps more relevant, that a zero entry in the matrix implies that no interaction occurred for this specific dyad.

In the analysis of dyadic interaction, the model is necessarily componential, where the interaction consists of three major components: a main effect of the giver (giver effect), a main effect of the receiver (receiver effect), and the relation of the two individuals that is independent of the giver and receiver effects, referred to as the dyad (Back and Kenny, 2010; Kenny et al., 2006).

## 2.2. Experimental data analysis: attacking time in group-housed pigs

### 2.2.1 Experiment setup

In this study, the experimental data was collected from 797 Yorkshire pigs (409 gilts and 388 barrows) that were strategically mixed into 59 single-sex social groups and housed in grow-finish pens with 10-15 pigs per pen. In terms of prior social acquaintances, each social group included pairs or trios of animals that had shared a common nursery pen for seven weeks immediately before moving into the grow-finish pens. Prior social acquaintances also existed for some animals that had shared the same litter after farrowing (10 weeks before mixed into the grow-finish pens; these pigs were previously housed together as a litter before weaning). No prior social acquaintance was assumed to exist for animals that were housed together for the first time after being mixed into the grow-finish pens. At the beginning of the experiment the average weight of the animals was 27.09 kg (SD±4.07). The experiment has been described in detail in previous studies (Angarita et al., 2019; Wurtz et al., 2017).

Pigs were video recorded five hours after mixing and four hours on the following morning (no overnight recording was performed). Videos were decoded manually by trained observers who recorded all attacks, their duration, and the identity of giver and receiver. After decoding, the total amount of time for each dyadic interaction was computed as described by Angarita et al. (2019). The directional aggression duration $y_{ijk}$ was defined as the total time in seconds that animal $i$ spent attacking another group mate animal $j$ within social group $k$ during the 9-hour post-mixing period. The final dataset contained 10,032 records consisting of total attacking duration for all possible dyads. Among those records, 1,100 pairs of animals (2,200 records) shared the same nursery pen prior to being remixed into the grow-finish pens, and 367 pairs of animals (734 records) were from the same litter.

### 2.2.2 Analysis model

After extensive model assessment and comparison, a hurdle Bernoulli-lognormal model was adopted. To keep things simple, in this paper it was assumed that a positive continuous response could be adequately modeled using a lognormal distribution and that a Bernoulli distribution could model the response when it is zero. However, the general principles presented here can be easily extended to other types of distributions as mentioned in the discussion. Thus, in this application, there are two sub-models (Equation 1). One sub-model estimates the probability of observing a zero (no attacks) while the other sub-model represents the duration of attacks conditional on its occurrence:

$$\begin{cases} y_{ijk} \sim Bernoulli(\theta_{ijk}), & if\ y_{ijk} = 0 \\ y_{ijk} \sim Lognormal(\mu_{ijk}, \sigma^2), & if\ y_{ijk} > 0 \end{cases} \quad [1]$$

where, $y_{ijk}$ is the total duration of the behavioral interactions between animal $i$ and animal $j$ in social group $k$ (in $y_{ijk}$, the first subindex corresponds to the aggression giver, the second subindex corresponds to the aggression receiver, and the third subindex indicates the social group), $\theta_{ijk}$ is the expected probability of the total attacking duration being zero for animals $i$ and $j$ in social group $k$. Further, $\mu_{ijk}$ is the mean of natural logarithm of $y_{ijk}$, and $\sigma^2$ is the variance of natural logarithm of $y_{ijk}$.

The transformed $\theta_{ijk}$ and $\mu_{ijk}$ have linear relationships with the explanatory variables (Equation 2):

$$\begin{cases} \log\left(\frac{\theta_{ijk}}{1-\theta_{ijk}}\right) = \mu'_{ijk} = FE'_{ijk} + g'_i + r'_j + d'_{ij} + sg'_k, & if\ y_{ijk} = 0 \\ \mu_{ijk} = FE_{ijk} + g_i + r_j + d_{ij} + sg_k, & if\ y_{ijk} > 0 \end{cases} \quad [2]$$

where $\mu'_{ijk}$ and $\mu_{ijk}$ are expected values (mean) on an underlying linked scale that can be modeled as linear combinations of individual-level and dyad-level systematic effects (described below). Note that notations with the superscript represent effects to model the probability of presenting no attacks while effects without the superscript model the attacking duration if the attack occurred. $d'_{ijk} \sim N(0, \sigma'^2_d)$ and $d_{ijk} \sim N(0, \sigma_d^2)$ represent random dyad effects. $sg'_k \sim N(0, \sigma'^2_{sg})$ and $sg_k \sim N(0, \sigma_{sg}^2)$ are random social group effects. The parameters $g'_i$, $g_i$, $r'_j$, and $r_j$ are explained in the following section. Further, the fixed effects (overall means) $FE'_{ijk}$ and $FE_{ijk}$ in Equation 2 are defined as:

$$\begin{cases} FE'_{ijk} = sex'_k + \alpha' x_{jk} + \beta' w_{ik} + \delta'_1 z^1_{ijk} + \delta'_2 z^2_{ijk}, & if\ y_{ijk} = 0 \\ FE_{ijk} = sex_k + \alpha x_{jk} + \beta w_{ik} + \delta_1 z^1_{ijk} + \delta_2 z^2_{ijk}, & if\ y_{ijk} > 0 \end{cases} \quad [3]$$

where $sex'_k$ and $sex_k$ are sex effects in social group $k$, $x_{jk}$ is the (within-group) centered weight of the receiver animal $j$ from social group $k$, $w_{ik}$ indicates the (within-group) centered weight of the giver animal $i$ from social group $k$, $z^1_{ijk}$ represents whether animal $i$ and animal $j$ from social group $k$ shared the same nursery group previously ($z^1_{ijk} = 0$ if they did not; otherwise, $z^1_{ijk} = 1$), and $z^2_{ijk}$ indicates whether animal $i$ and animal $j$ from social group $k$ were previously housed together in the same litter before weaning ($z^2_{ijk} = 0$ if they were not; otherwise, $z^2_{ijk} = 1$). Finally, $\alpha'$, $\alpha$, $\beta'$, $\beta$, $\delta'_1$, $\delta_1$, $\delta'_2$, and $\delta_2$ denote the corresponding coefficients of the exploratory variables. Without losing generality, we illustrate a linear model where the response can be simply decomposed into giver effects, receiver effects, and dyad-specific effects in Figure 2.

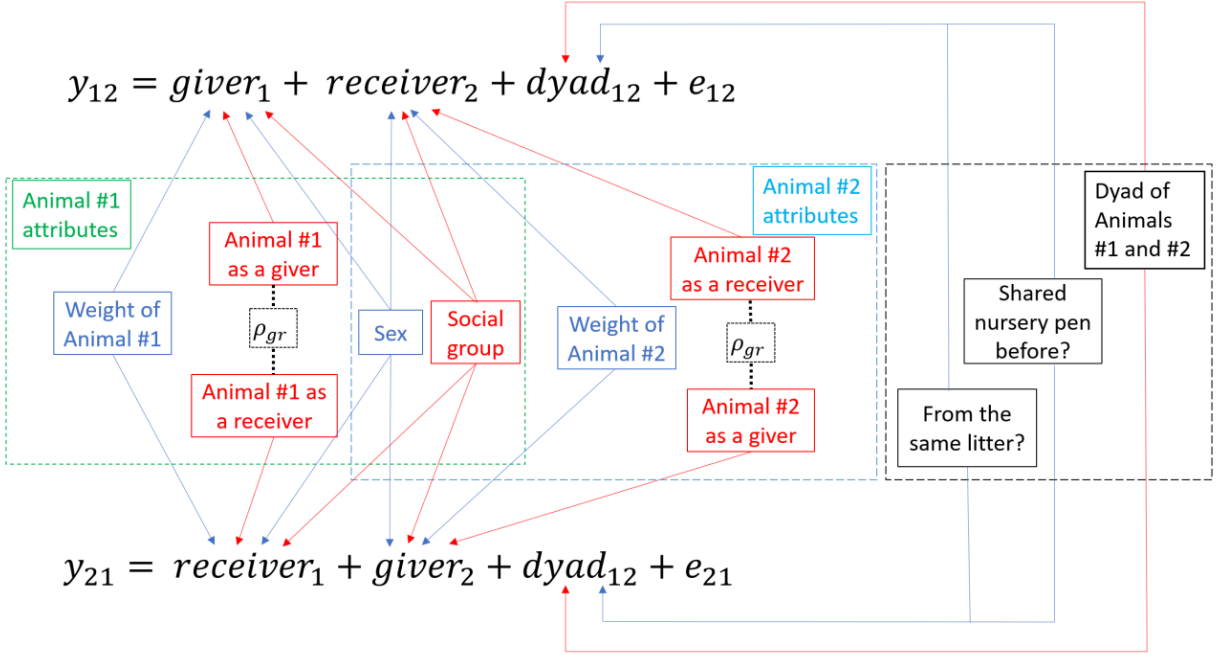

**Figure 2** Illustration of a dyadic interaction model as an example that partitions the response into giver effects, receiver effects, and dyad-specific effects. Blue lines/arrows mean fixed effects, and red represents random effects, e stands for the residual term.

**2.2.3 Modeling of (co)variances**

Under the model in Equation 2, effects of the giver and the receiver are modeled for each animal. Those two effects covary, assuming:

$$\begin{pmatrix} g'_j \\ r'_j \end{pmatrix} \sim N\left( \begin{pmatrix} 0 \\ 0 \end{pmatrix}, \begin{pmatrix} \sigma'^2_g & \sigma'_{gr} \\ \sigma'_{gr} & \sigma'^2_r \end{pmatrix} \right) \quad [4]$$

$$\begin{pmatrix} g_j \\ r_j \end{pmatrix} \sim N\left( \begin{pmatrix} 0 \\ 0 \end{pmatrix}, \begin{pmatrix} \sigma^2_g & \sigma_{gr} \\ \sigma_{gr} & \sigma^2_r \end{pmatrix} \right) \quad [5]$$

where $\sigma'_{gr}$ and $\sigma_{gr}$ represent the covariance between the receiver and the giver effects of the same animal. Moreover, $\sigma'_{gr}$ or $\sigma_{gr}$ could take negative values. For example, $\sigma_{gr} < 0$ if an animal spends more time attacking other animals but receives less aggression (in terms of duration) from other animals. Contrarily, $\sigma_{gr}$ will be positive in those cases where animals that deliver more aggression also receive more aggression from other animals. A similar analysis could be done for $\sigma'_{gr}$ but related to the probability of not delivering attacks. The magnitude and sign of these parameters are of importance to behaviorists.

We also derive the estimated giver-receiver correlation as this is easily interpretable to applied scientists:

$$\rho'_{gr} = \frac{\sigma'_{gr}}{\sigma'_g \sigma'_r}, \quad \rho_{gr} = \frac{\sigma_{gr}}{\sigma_g \sigma_r} \quad [6]$$

The relative magnitude of $\sigma'^2_g, \sigma'^2_r, \sigma^2_g,$ and $\sigma^2_r$ are also important. A relatively small value for a specific source of variation means that the process is mostly driven by other random sources.

**2.2.4 Estimation**

For statistical analysis, the model represented in Equations 1-3 could be fitted using restricted maximum likelihood or using Bayesian methods. We chose to use a Bayesian approach (Box and Tiao, 2011). Details for the implementation of model fitting are provided in the appendix. In the companion GitHub (https://github.com/jun-jieh/DyadAnalysis) we provide examples for implementation. A total of 4,000 Markov chain Monte Carlo (MCMC) samples were generated for parameter estimates. The parameter $\theta$ for a fixed effect given the observed data $y$ was considered significant ($P<0.05$) if

$$1 - max\left(p(\theta < 0|y), p(\theta > 0|y)\right) < 0.05 \quad [7]$$

where $p(\theta < 0|y)$ means that the probability of the parameter $\theta$ is smaller than zero while $p(\theta > 0|y)$ represents the probability of $\theta$ being larger than zero, and the function *max*() returns the maximum value of the two elements. In practice, these probabilities are estimated based on the relative frequencies obtained from the MCMC samples.

**2.2.5 Validation strategies and posterior predictive checks: how well does the model fit the data?**

Posterior predictive checking is an important part of model evaluation. For this checking, new data are simulated conditional on the fitted model and their distribution is compared to observed data (Gabry et al., 2019). Moreover, this posterior predictive checking can be done with internal validation (all the data are used for model fitting and for validation) or using external validations (also known as out-of-sample or hold-out validations) (Vehtari et al., 2017; Vehtari and Ojanen, 2012), where the data are split into a training set (used for model fitting) and a validation set (used for the validation/checking). In the case of external validation, the way data are split is very important. Specifically, we split the entire dataset into a training set and a validation set using three different strategies:

1. A stratified 5-fold cross-validation (Vehtari and Ojanen, 2012) was used where in each fold, a random subset of each social group (80% of the data) was utilized for model training, while the remaining records were for testing purpose. It maintains the same social group ratio throughout the five folds as the ratio in the original (entire) dataset.
2. A block-by-social-group (5-fold) cross-validation was performed. In each fold, all records from randomly selected social groups that make up approximately 80% of the entire data were pooled and used for training purpose, while the remaining (validation data comprising 20% of the observations) set was from the left-out social groups that were not part of the training data.
3. A block-by-focal-animals validation was proposed and run for five replicates. In this validation scenario, we selected seven animals from each group and used all their aggression records as both aggression givers and aggression receivers for the training set. The validation set contained only interactions between non-focal animals. This resembles a common way in which videos could be decoded; only some animals are followed and all their interactions with everyone else are decoded. Furthermore, by selecting seven focal animals per group, the resulting training set size was 78% of the entire data, while the remaining data (approximately 22% of all records) was used for testing. This led to a similar set size for comparison to the other two model checking strategies.

In each validation strategy, the model was fitted five times (for the five folds or replicates), and as part of the Bayesian model fit procedure, 500 MCMC samples were generated from the posterior predictive density. The posterior distribution of the generated samples was compared to the distribution of the observed dataset, where the response variables were transformed into a logarithm scale. To evaluate the predictive performance, Pearson correlation and root mean square error (RMSE) (Chai and Draxler, 2014) between the log-transformed (observed) response i.e. log(response+1) and the mean of log-transformed predicted response (linear predictors in Equation 2) in the validation set was computed across all validation scenarios. In addition, area under the ROC (Receiver Operating Characteristics) curve (Ling et al., 2003), also known as AUC, was computed to evaluate performance on the prediction of attack presence/absence .

## 2.3 Ethical approval

All animal protocols were approved by the Institutional Animal Care and Use Committee (Animal Use Form number 01/14-003-00).

## 3. Results

### 3.1 Estimation of animal-specific effects, dyad-specific effects, and (co)variance components

Table 1 shows the posterior distributions of the individual animal effects and dyadic effects. The random dyad effect was not estimable (the model including the dyad effect did not converge; see appendix for details).The nursery mate experience (animals in a dyad knew each other from sharing the same nursery pen prior to being remixed into grow-finish pens) exhibited significant effects, reducing the probability of presenting attacks and, if attacks happened, reducing duration of them ($\delta'_1 = 0.505, P < 0.05$; and $\delta_1 = -0.501, P < 0.05$; Table 1). The estimates indicated that for the dyads where the pigs had nursery mate experience, we would expect to see 65.7% increase of the odds of presenting no attacks; on the other hand, if the dyad did present attacks, pigs with nursery mate experience exhibited a 39.4% decrease in attacking duration. This means that if animal $i$ and animal $j$ were housed in the same nursery pen previously and then are remixed into a grow-finish pen, as might be expected under production conditions, they are less likely to attack each other and if they do attack each other, the length of attacking duration will be significantly shorter than the average attacking time of two animals who had not recently been housed together. The remaining animal-specific properties (weight of giver, weight of receiver, and sex) and dyad-specific attributes (whether the giver and the receiver were from the same litter) were not significant.

**Table 1.** Estimated posterior statistics for fixed effects and (co)variance components explained on total attacking duration between the giver animal and the receiver animal. Q: quantile.

| Parameter | | $y_{ijk} = 0$ | | | | $y_{ijk} > 0$ | | | |
|---|---|---|---|---|---|---|---|---|---|
| | | Mean | Q 2.5% | Q 50% | Q 97.5% | Mean | Q 2.5% | Q 50% | Q 97.5% |
| $sex', sex$ | | -0.083 | -0.483 | -0.085 | 0.312 | -0.039 | -0.162 | -0.038 | 0.085 |
| $\alpha', \alpha$ | Receiver weight | 0.001 | -0.011 | 0.001 | 0.013 | -0.001 | -0.008 | -0.001 | 0.006 |
| $\beta', \beta$ | Giver weight | -0.008 | -0.027 | -0.008 | 0.012 | -0.007 | -0.017 | -0.007 | 0.003 |
| $\boldsymbol{\delta'_1, \delta_1}$ | Nursery mate | **0.505** | **0.392** | **0.504** | **0.618** | **-0.501** | **-0.573** | **-0.501** | **-0.424** |
| $\delta'_2, \delta_2$ | Litter mate | -0.170 | -0.357 | -0.171 | 0.015 | -0.023 | -0.139 | -0.023 | 0.087 |
| $\sigma'^2_g, \sigma^2_g$ | Giver variance | 1.050 | 0.737 | 1.035 | 1.443 | 0.063 | 0.044 | 0.062 | 0.089 |
| $\sigma'^2_r, \sigma^2_r$ | Receiver variance | 0.023 | 0.008 | 0.021 | 0.045 | 0.002 | 0.001 | 0.002 | 0.005 |
| $\rho'_{gr}, \boldsymbol{\rho_{gr}}$ | Giver-receiver correlation | 0.155 | -0.015 | 0.155 | 0.328 | **0.203** | **0.009** | **0.202** | **0.397** |
| $\sigma'^2_{sg}, \sigma^2_{sg}$ | Social group variance | 0.473 | 0.282 | 0.460 | 0.735 | 0.020 | 0.001 | 0.019 | 0.052 |
| $\sigma^2$ | Error variance | - | - | - | - | 1.076 | 1.031 | 1.075 | 1.123 |

The giver-receiver correlation was not significant for the Bernoulli sub-model (when estimating the probability of animal $i$ not presenting attacks to animal $j$; Table 1). On the other hand, a weak but significant correlation was obtained in the lognormal sub-model to analyze the attacking duration ($\rho_{gr} = 0.203, P < 0.05$). This means that when one pig spends more time delivering aggression this same pig will also receive longer attacks.

**3.2 Predictive performance in different validation strategies**

We assessed the fitted model through posterior predictive model checks by inspecting two important aspects of: 1) how well it predicted the probability of not having an attack, and 2) how well it predicted the mean duration of the attacks when they occur.

Figure 3 presents the posterior predictive distribution of the probability of observing no attacks between animals (relative frequency of zeros). The distribution of the proportion of predicted zeros across multiple replicates of simulated data (lighter bins in Figure 3) was centered around the proportion of zeros in the observed response (dark solid lines in Figure 3). That is, regardless of training-validation data partitions, the estimated validation proportion fell well within the posterior predictive density in all validation strategies.

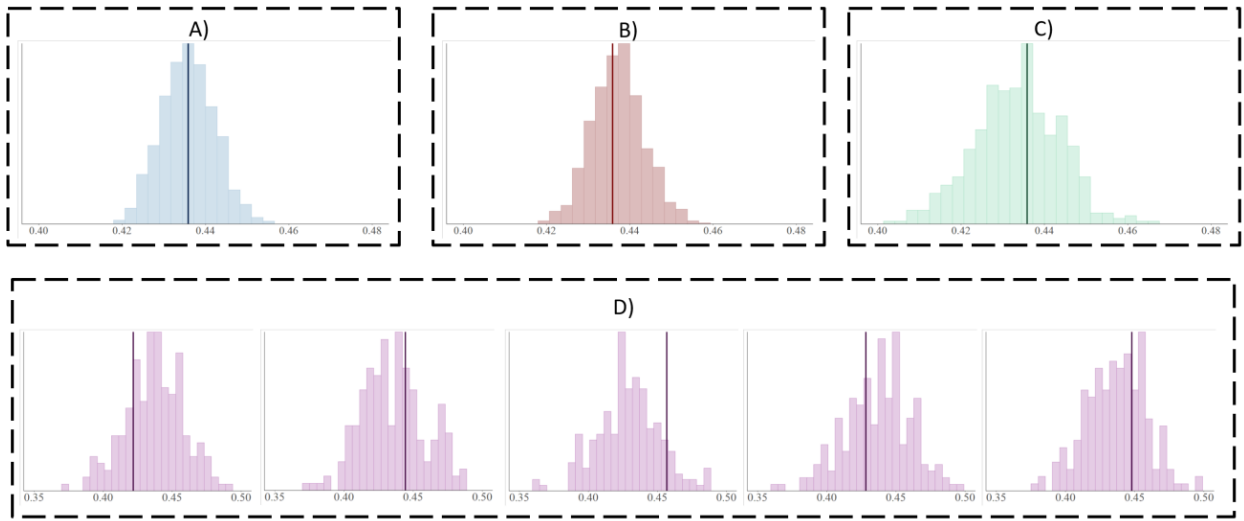

**Figure 3** Proportion of zeros of validation set y (dark lines), with proportions of zeros for 500 simulated datasets ỹ drawn from the posterior predictive distribution (lighter bins). A), the model that used all data points for model fitting to predict the same dataset; B) stratified 5-fold cross-validation; C) Block-by-social-group cross-validation; D) 5 replicates of block-by-focal-animals validation.

For each of the validation strategies, Figure 4 shows the distribution for the means of the simulated data (light bins) and the observed data (dark solid line). The response variables were transformed into a logarithm scale i.e., log(response+1). In general, the simulated data were consistent with the observed data (no systematic lack-of-fit was observed). However, in the internal validation and block-by-focal-animals validation, the mean duration of attacks was better approximated compared to when the stratified 5-fold and block-by-social-group cross-validation approaches were used.

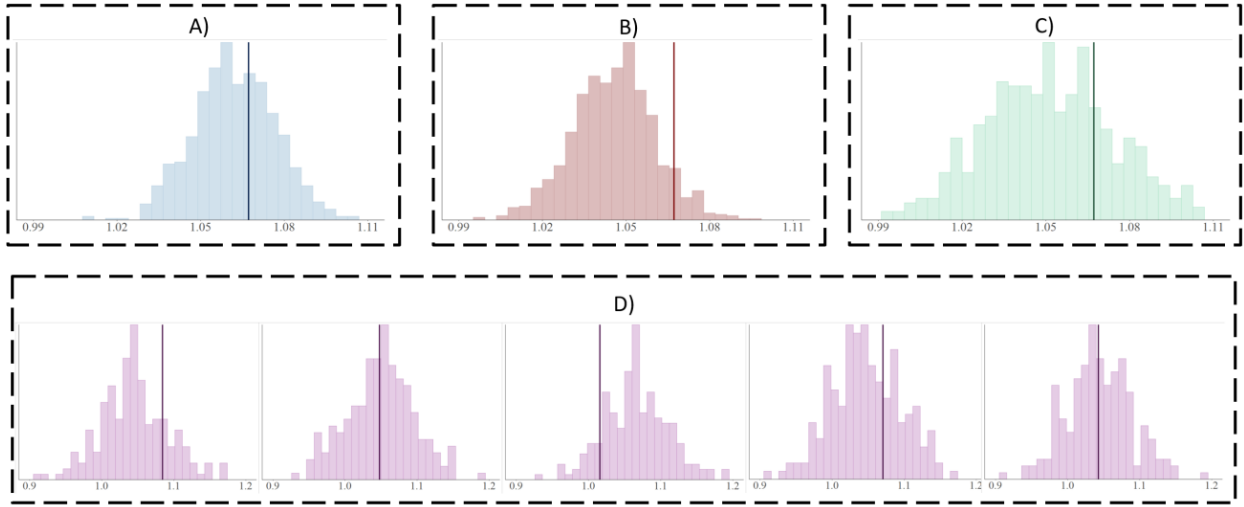

**Figure 4** Distribution for the mean value of all observations across replicates. Mean of the validation set y (dark solid line) is compared with the means of 500 simulated datasets ỹ drawn from the posterior predictive distribution (lighter bins). We compared the logarithm of the observed and the simulated variables i.e. log(y+1) and log(ỹ +1). A), the model that used all data points for model fitting to predict the same dataset; B) stratified 5-fold cross-validation; C) Block-by-social-group cross-validation; D) 5 replicates of block-by-focal-animals validation.

Correlation and RMSE of the log-transformed response and AUC for the prediction of presence/absence of attacks in the validation set were computed across all validation scenarios (Table 2). The metrics allow for comparison between different validation strategies. Compared to the internal validation (all the data were used for model fitting and for validation), the predictive performance of the stratified 5-fold cross-validation, block-by-social-group cross-validation, and block-by-focal-animals validation showed much lower correlation and AUC, and larger RMSE (Table 2). Notably, the predictive performance in block-by-social-group validation was consistently worse than the remaining ones.

**Table 2.** Metrics for evaluating predictive performance of the model under different validation strategies. AUC, area under ROC (Receiver Operating Characteristics) curve; RMSE, root mean square error; CV, cross-validation.

|  | Pearson correlation | AUC | RMSE |
| --- | --- | --- | --- |
| In-sample validation | 0.595 | 0.826 | 1.173 |
| Stratified 5-fold CV | 0.228 (SD±0.014) | 0.653 (SD±0.017) | 1.391 (SD±0.020) |
| Block-by-social-group CV | 0.115 (SD±0.023) | 0.523 (SD±0.017) | 1.422 (SD±0.023) |
| Block-by-focal-animals validation | 0.286 (SD±0.020) | 0.532 (SD±0.013) | 1.362 (SD±0.014) |

## 4. Discussion

In this study, we have illustrated how to use GLMMs to analyze dyadic data from animal behavior studies that record interactions between animals in social groups. Through changing distributional assumptions and link functions, this approach can be easily adapted to analyses of categorical, ordinal, count, and continuous response types. Instead of modeling an individual animal's response, the proposed model exhibits advantages of analyzing interactive behaviors of pairs of animals in terms of flexibility and interpretability. Furthermore, the inclusion of random and fixed effects specific to each giver, receiver, and dyad (when possible) contributes to partitioning the observed variance into interpretable components.

Several approaches have been used in the analysis of animal behavioral interactions. A commonly used approach ignores the dyadic nature of the data and sums over rows or columns of an interaction matrix to simply obtain total time spent by each individual engaged in the behavior of interest (Figure 1a). Following this summation, linear models are used to study several sources of individual-level effects on the behavior of interest. We call this a 'marginal analysis' as it operates on the margin of the interaction matrix. For example, Savory and Mann (1997) studied the effects of genetic strain, age, and feeding pattern on aggressive pecking behavior of pullets, where for each individual the proportion of the aggressive behavior was computed (i.e., the total time of aggression was summed and divided by the length of observation period). Similarly, in two other studies (Turner et al., 2009, 2008), the authors recorded and treated the total duration of nonreciprocal aggression delivered and received by each individual pig as response variables, and they fitted linear mixed models to estimate additive genetic effects on the marginal response. In addition, Verdon et al. (2018) studied aggressive behavior of sows where the unit of analysis was a group of sows. They counted the frequency of aggressive interactions from all possible pairs of animals within each group and fitted generalized mixed models to analyze the marginal response. A shortcoming of these analyses is that the effect of dyadic factors cannot be investigated. The proposed approach of this study allows the inclusion of both animal-level effects (marginal effects) as well as dyadic effects relevant to that particular pair of animals. For instance, we can add previous group- or littermate experience into the model for each dyad. In addition, genetics/genomics information of the giver and the receiver, as well as their genetic relationship, can be further included as an extensive form of our proposed model.

Another common approach analyzes dyadic interactions as independent observations. Oldham et al. (2020) fitted a linear mixed model to investigate effects of characteristics of both pigs on the initiating pig's latency to initiate agonistic behavior in a dyadic contest. This study was carefully designed and analyzed such that only one observation per animal and per contest (dyad) was available. This allowed the use of a simple linear model for the analysis. However, more precisely, dyads refer to relation of two individuals embedded in a social context (Kenny et al., 2020), while Oldham et al. (2020) manually selected paired pigs for contests instead of selecting dyads from a social context. In dyadic data extracted from multiple social groups with more than two individuals per group where each pig is exposed to multiple group mates, the assumption of independence between observations does hold i.e., the dyad is the fundamental unit of analysis (Kenny et al., 2020), and the proposed approach allows modeling the variances and co-variances of social groups, dyads, and individuals in a very straightforward way. For instance: we include the giver and receiver effects and account for their correlation. Thus, our proposed model is particularly useful for studying social interactions where animals are housed in multiple social groups over time.

In addition to introducing a model for the analysis of dyadic data in studies of social animal behavior, this study yields valuable results for understanding factors that affect post-mixing aggression in growing pigs.

The results indicate that the giver explained more variation of the dyadic interaction than the receiver (Table 1). To the best of our knowledge, only one previous publication has used a GLMM to dissect the giver and receiver effects in animal behavior data (Wang et al 2022), however, they did not consider the inclusion of dyadic fixed effects or the inclusion of a dyad-level random effect. A related line of research used bivariate marginal models to study delivery and reception of non-reciprocal aggression (Turner et al., 2009, 2008). Interestingly, both studies found that delivery of aggression was more heritable than reception of aggression. This encourages further analyses with the dyadic model to tease apart genetic effects from environmental effects.

One application in human behavioral ecology (Koster et al., 2015) proposed using GLMMs to perform dyadic analysis of food sharing between households and reported that the meal giver explained 75% of the variance components while the variance ratio of the meal receiver was 6%, which showed results quantitatively similar to ours. Given more complete datasets (with more observed variables of the interacting individuals), ethologists could further use the proposed dyadic model to dissect factors that may influence delivery of the behavior as well as the characteristics of the receiver that attract the behavior.

In the context of post-mixing aggression in finishing pigs, we estimated the correlation between the random giver effect and the random receiver effect of the same individuals (Table 1). In a previously published marginal analysis of post-mixing aggression in pigs, the correlation between delivering and receiving non-reciprocal aggression did not differ significantly from zero (Turner et al., 2008). However, our model revealed a weak but significant correlation between the giver and receiver effects on the duration of the attacks. This means that animals which attack for longer duration also receive longer attacks themselves, and this could be a result of receiver animals defending themselves and striking back (i.e., receivers may use attacks as a form of defense) (Oldham et al., 2020). The aggregated data used in this study did not allow investigation of the sequences of attacks (as our dyadic data was defined as the total aggression duration from a giver to a receiver), thus further work is needed to analyze heterogeneous and repeated measures of dyadic interactions over time.

Interestingly, our model did not yield a significant effect of the bodyweight of giver or receiver on the occurrence or duration of attacks. It is worth mentioning that the goal of this study was not to investigate bodyweight effects of the giver and receiver on attacking duration. Our result for the bodyweight effect might be due to the limited variation in body size within the social groups of our study as we had deliberately mixed together animals of similar body size in the finishing groups. This could result in a non-significant effect of the animal weight given limited variation in those covariates. The literature on this matter (bodyweight effect) does not offer a definite conclusion regarding the effect of bodyweight on aggressive behaviors of pigs. In one study of aggressive contests between pigs (Oldham et al., 2020), neither the weight of the contest initiating pig nor the weight difference between the contestants significantly influenced the latency to initiate the aggression. This agrees with our findings on bodyweight effects. However, in another study of dyadic contests in pigs, the winner pigs were significantly heavier than the loser pigs (Camerlink et al., 2019). We need to point out that initiating an attack and winning a contest are different. Our result suggests that pigs might not be good at telling whether they were going to win or not when they decided to attack. Camerlink et al. (2015) also showed that between pairs of size-matched pigs, pigs which were more likely to be attackers were not more likely to be winners.

The variance component of random dyad effect (the effect of giver-receiver relation; see Sections 2.1.2 and 2.2.2) was not estimable in this study; however, we found that having shared nursery pens immediately before being mixed into grow-finish pens (a dyad-level covariate) showed a significant effect ($P<0.05$; Table 1). This finding is confirmed in the literature, for instance, Li and Wang (2011) reported that unfamiliar pigs fought for longer durations and fought more frequently than familiar pigs, when pigs were remixed into new social groups. However, another dyadic level predictor (whether the two pigs were previously housed together as a litter before weaning; the pigs were housed as a litter for approximately three weeks before introduced to nursery pens) was not significantly associated with delivery or duration of aggression. This hints at the fact that animals who once shared a social group several weeks prior to the mixing, even if they are related, are unlikely to remember each other. It is unclear how long pigs remember each other though a possible time range could be three to six weeks (Mendl et al., 2010). Since pigs in this study spent approximately seven weeks in nursery pens immediately before being remixed into grow-finish pens, it was possible that pigs did not recognize their initial littermates when re-introduced to them in grow-finish pens.

In addition to the models presented in this study, we also evaluated other GLMMs, including log-Poisson, zero-inflated log-Poisson, Gaussian, and zero-inflated Gaussian. The posterior predictive checks conducted (results not shown) showed that the hurdle-Bernoulli model was the one that fitted the data better. The hurdle model did so by dissecting the trait into two components: the tendency of not delivering attacks and a second component of the attacking duration. The complexity of the model may limit its practical use as there are two correlated traits rather than one per animal that can be used for decision making. However, it is worth mentioning that other GLMMs may be adequate for other settings. In the companion GitHub (https://github.com/jun-jieh/DyadAnalysis), we provide simpler GLMMs that are more general and can be easily adapted. In addition, to check model fitting, the in-sample posterior predictive checking (predicting the data used for model fitting) suffices, but for studying the model's ability to predict future data, out-of-sample validation (predicting observations left out of the model fitting process) should be used.

Social interaction data has been recently used as predictors of other traits. For example, Turner et al. (2020) constructed play fighting social networks of pigs using dyadic interaction data and extracted individual level and network level traits to build prediction models for lesion score counts. In a different application, Angarita et al. (2019) proposed using the dyadic matrices of aggression duration between pigs to parametrize social genetic effects of lesion scores. In these cases, the dyadic data (or its derived social network features) were used as a predictor rather than as a response variable. Nevertheless, the proposed predictive modeling of dyadic data could be used to add uncertainty to these applications. For instance, the internal and external validation (see Section 2.2.5 and Figures 3 and 4) used for model checking provides a natural way to resample plausible social interaction matrices that could be then subject to social network analysis or included in social genetic effects modeling. Moreover, obtaining the summation of all dyadic interaction of an animal as a giver (or receiver) allows for predicting individual-level aggressiveness (or vulnerability), for instance, the marginal intensity as shown in Figure 1a. Such individual level phenotypes can be used for management and as traits in genetic evaluations. Further experiments could be designed to validate the early prediction of animal social behavior that can be further related to animal welfare and production traits.

In this study, we considered several ways of splitting data for training (model fitting) and validation. Different validation scenarios (see Section 2.2.5) could be related to possible situations in real-life applications or relevant prediction problems (Burgueño et al., 2012). The stratified 5-fold cross-validation was designed for evaluation when the model was used for predicting unobserved (directional) social behaviors between two animals. The block-by-social-group validation mimicked a situation where the effects of giver, receiver, and dyad were evaluated in some social groups but not in others. Similarly, the block-by-focal-animals validation mimicked a situation when the giver, receiver, and dyad effects were modeled given records related to the focal animals but not for non-focal animals.

The predictive performance of the fitted models varied depending on the validation strategies (Table 2). The block-by-social-group cross-validation yielded the lowest correlation. This could be the result of not accounting for factors affecting social group composition. In fact, Samarakone and Gonyou (2009) have suggested that pigs may shift their aggressive behaviors accordingly to the composition of their social groups. Consequently, this could be revisited in further analyses and experimental setups where more group-specific variables are recorded and included in the dyadic model. In short, animal behavioral studies may consider introducing group-specific effects into the proposed model and exploring how these effects influence interactive behaviors.

The block-by-focal-animals validation yielded a slightly higher correlation and smaller RMSE compared to the stratified 5-fold cross-validation (Table 2). The result suggests that selecting focal animals and decoding their interactions with all other animals in the group may be a more efficient way to build predictive models of dyadic interactions than randomly selecting snippets of video for decoding. This idea has also been suggested by ethologists (Bosholn and Anciães, 2018). Furthermore, a dyadic model could be fitted using preliminary data to determine which factors better predict animal interactions, and then focal animals could be selected based on the significant factors to cover a large variation in responses. Such sampling strategies could be useful for improved manual video decoding efficiency.

## 5. Conclusion

We proposed an approach for the analysis of animals' social interactions based on modeling dyadic data. We illustrated its use through fitting a generalized linear model to total attacking time post-mixing between pairs of grow-finish pigs. Taking advantage of the flexibility and interpretability of the proposed model, we found that if two pigs had shared a common nursery pen immediately before being remixed into new social groups, they tended to spend less time engaging in the agonistic behavior. In addition, the positive correlation between the giver and receiver suggested that a pig that spent more time attacking was also more likely to be attacked for more time. The proposed model can be extended to incorporate additional giver-specific, receiver-specific, and dyad-specific effects. Moreover, we pursued alternative cross-validations and found that overlooking group-specific factors worsened the predictive performance of the proposed model. We also demonstrated that focusing on a fraction of all animals and decoding all their interactions with the remaining animals in the group is an effective way to perform inference and predictions on social interactions in the group while limiting the amount of time and effort dedicated to decoding video.

**Acknowledgements**

This work was funded by NIFA Awards 2017-67007-26176 and 2021-67021-34150.

**Competing interest statement**

We declare that we have no financial and personal relationships with other people or organizations which can inappropriately influence our work. There is no professional or other personal interest of any nature or kind in any product, service, and/or company that could be construed as influencing the position presented in, or the review of, the manuscript entitled.

**Appendix**

**1. Implementation detail of the Bayesian approach**

In parameter estimation, marginal posterior distributions of the parameters were obtained using Markov chain Monte Carlo method through Stan program. We ran four chains of 15,000 iterations, where we set the burn-in (warmup) to 5,000 iterations and every 10$^{th}$ samples were saved in each chain. Convergence diagnostics and graphical posterior predictive checks were performed using *rstan* and *bayesplot* packages in R (R Core Team, 2020). In the following parts of the appendix, we first present the prior distributions that were used in this study. We then show trace plots and autocorrelation plots of the fitted model. In addition, the summary of convergence diagnostics for the model with random dyad effect are included at the end. In the companion GitHub (https://github.com/jun-jieh/DyadAnalysis) we provide examples for implementations of multiple GLMMs and their model fitting that utilized a Bayesian method through *rstan* package in R (Carpenter et al., 2017; R Core Team, 2020).

**2. Prior distributions of model parameters that are described in Section 2.2**

$$\begin{pmatrix} \sigma'^2_g & \sigma'_{gr} \\ \sigma'_{gr} & \sigma'^2_r \end{pmatrix} \sim Wishart\left(4, \begin{pmatrix} 5 & 0 \\ 0 & 5 \end{pmatrix}\right)$$

$$\begin{pmatrix} \sigma^2_g & \sigma_{gr} \\ \sigma_{gr} & \sigma^2_r \end{pmatrix} \sim Wishart\left(4, \begin{pmatrix} 5 & 0 \\ 0 & 5 \end{pmatrix}\right)$$

$$sg'_k \sim Uniform(0,10)$$

$$sg_k \sim Uniform(0,10)$$

$$\sigma \sim Uniform(0,100)$$

i

$$sex'_k \sim Uniform(-10,10)$$

$$sex_k \sim Uniform(-10,10)$$

$$\alpha' \sim Uniform(-100,100)$$

$$\alpha \sim Uniform(-100,100)$$

$$\beta' \sim Uniform(-100,100)$$

$$\beta \sim Uniform(-100,100)$$

$$\delta'_1 \sim Uniform(-10,10)$$

$$\delta_1 \sim Uniform(-10,10)$$

$$\delta'_2 \sim Uniform(-10,10)$$

$$\delta_2 \sim Uniform(-10,10)$$

**3. Trace plots of posterior estimates of effects and variance components**

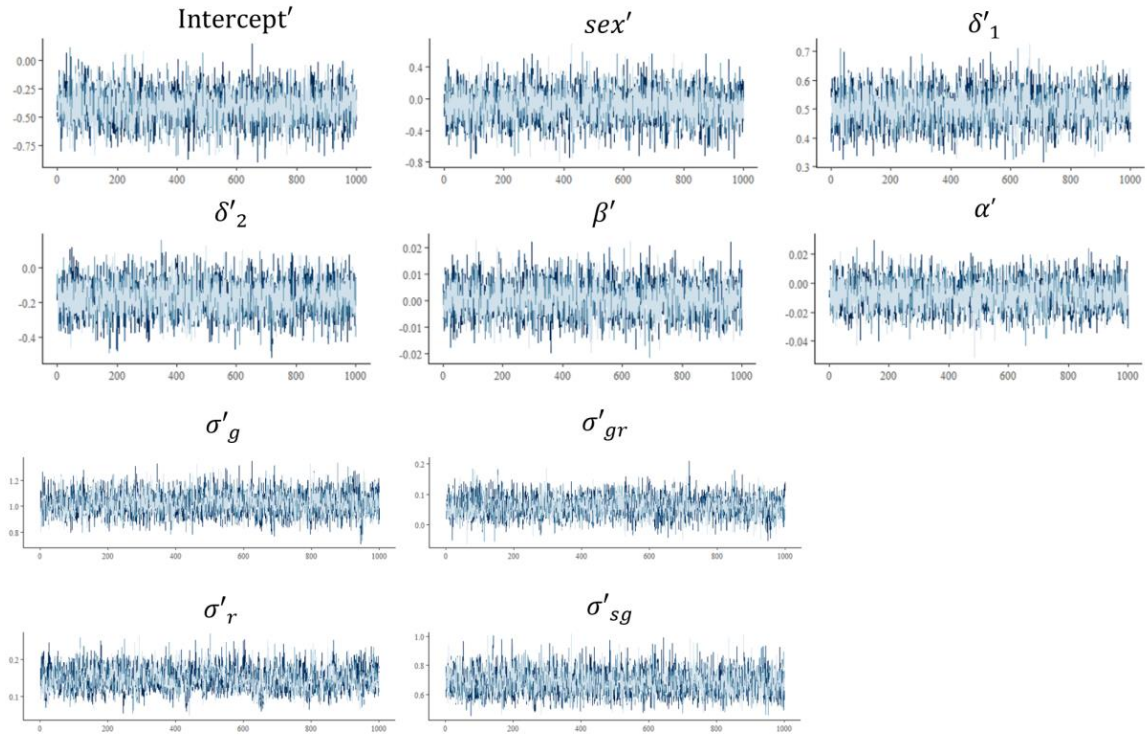
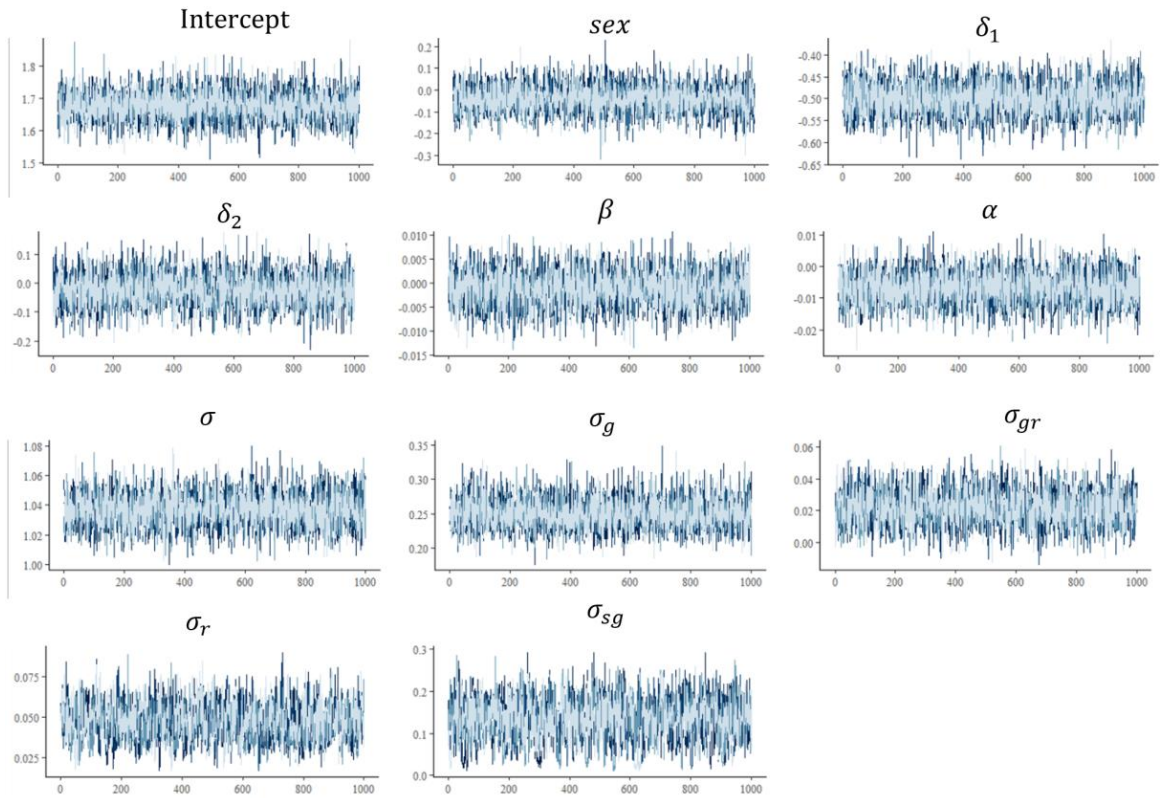

## 4. Autocorrelation plots by chain and by parameters

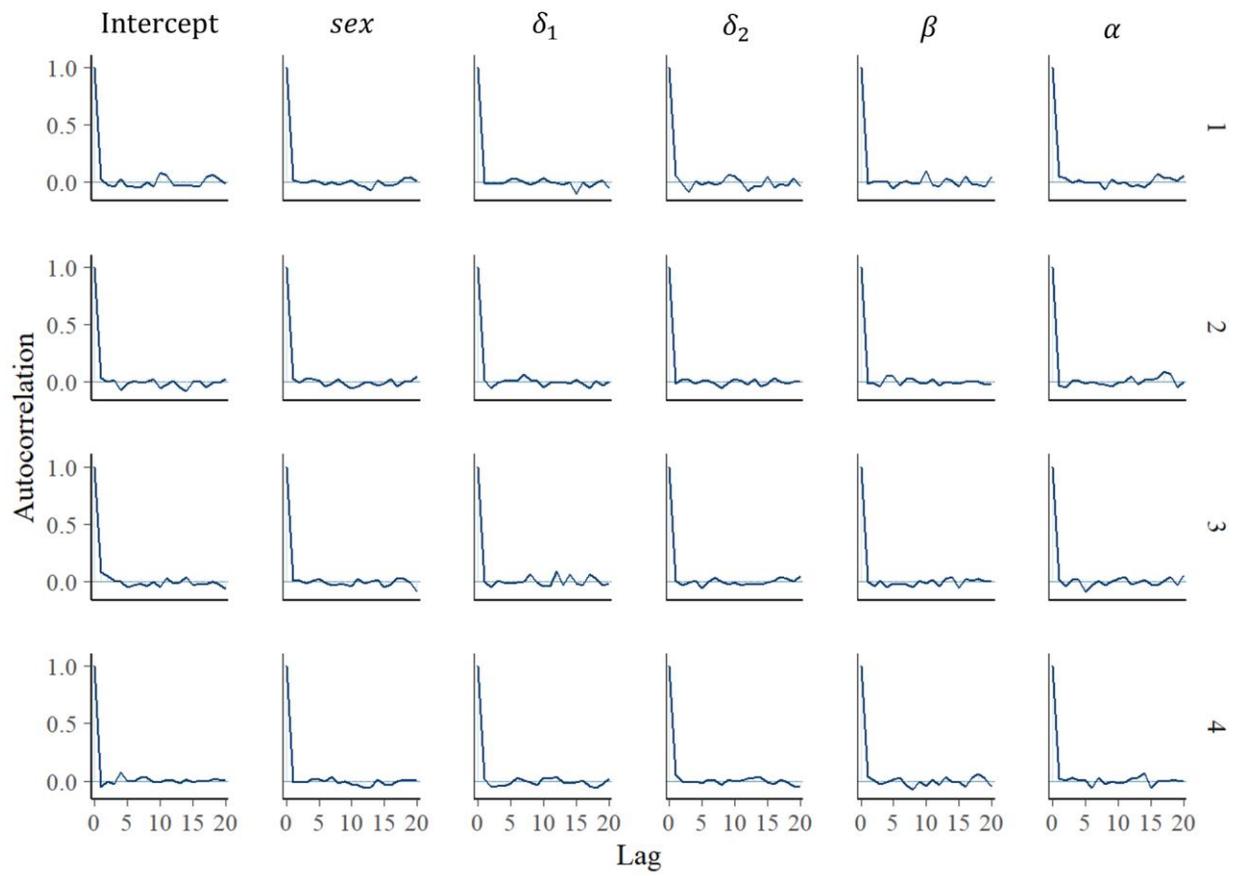

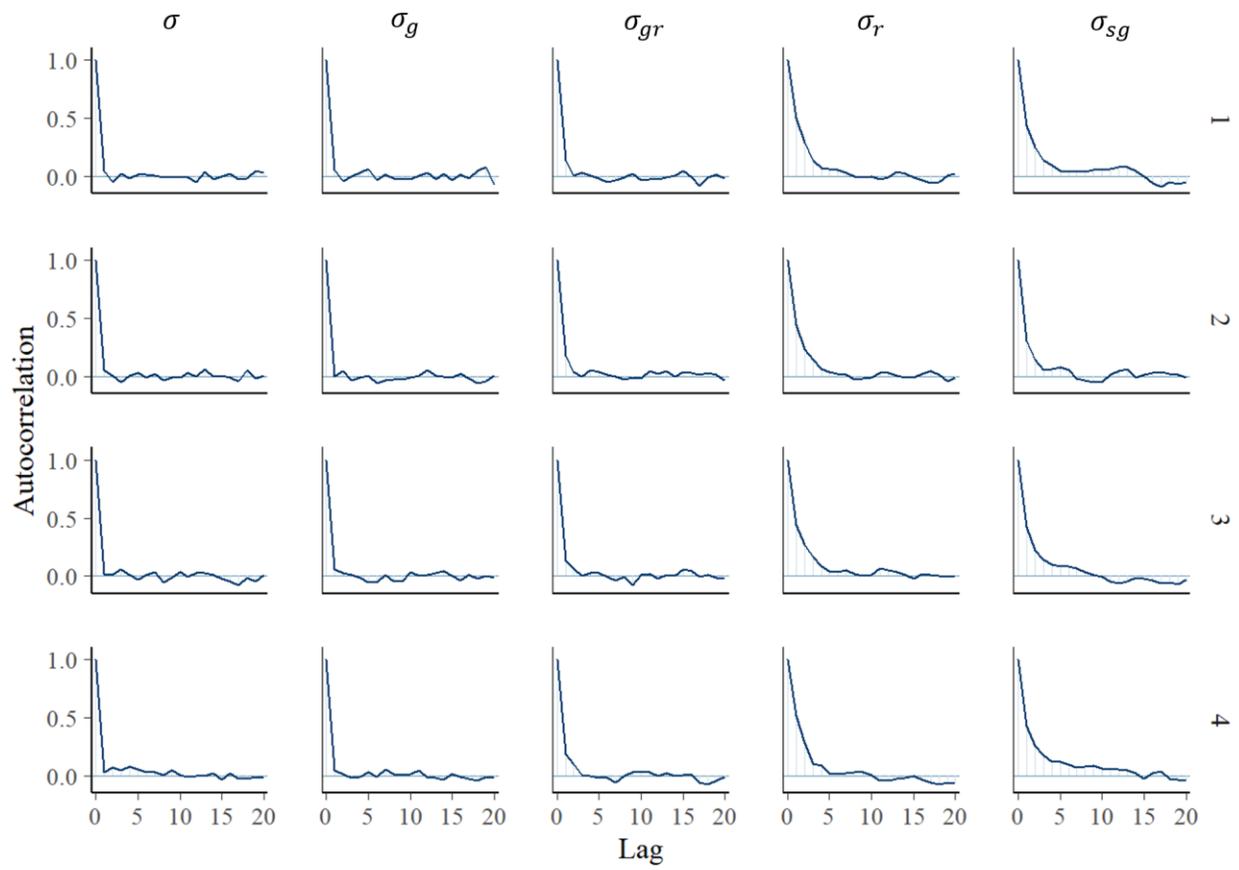

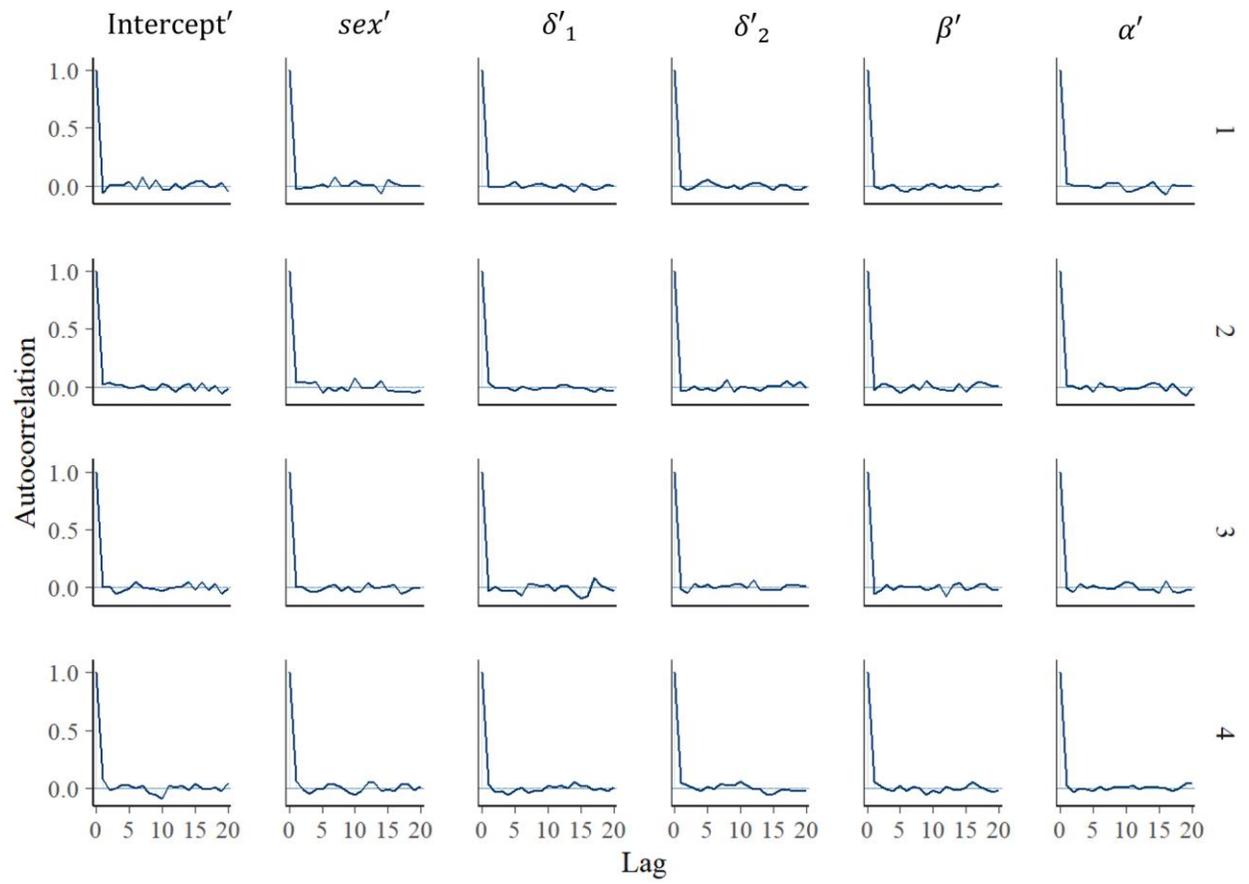

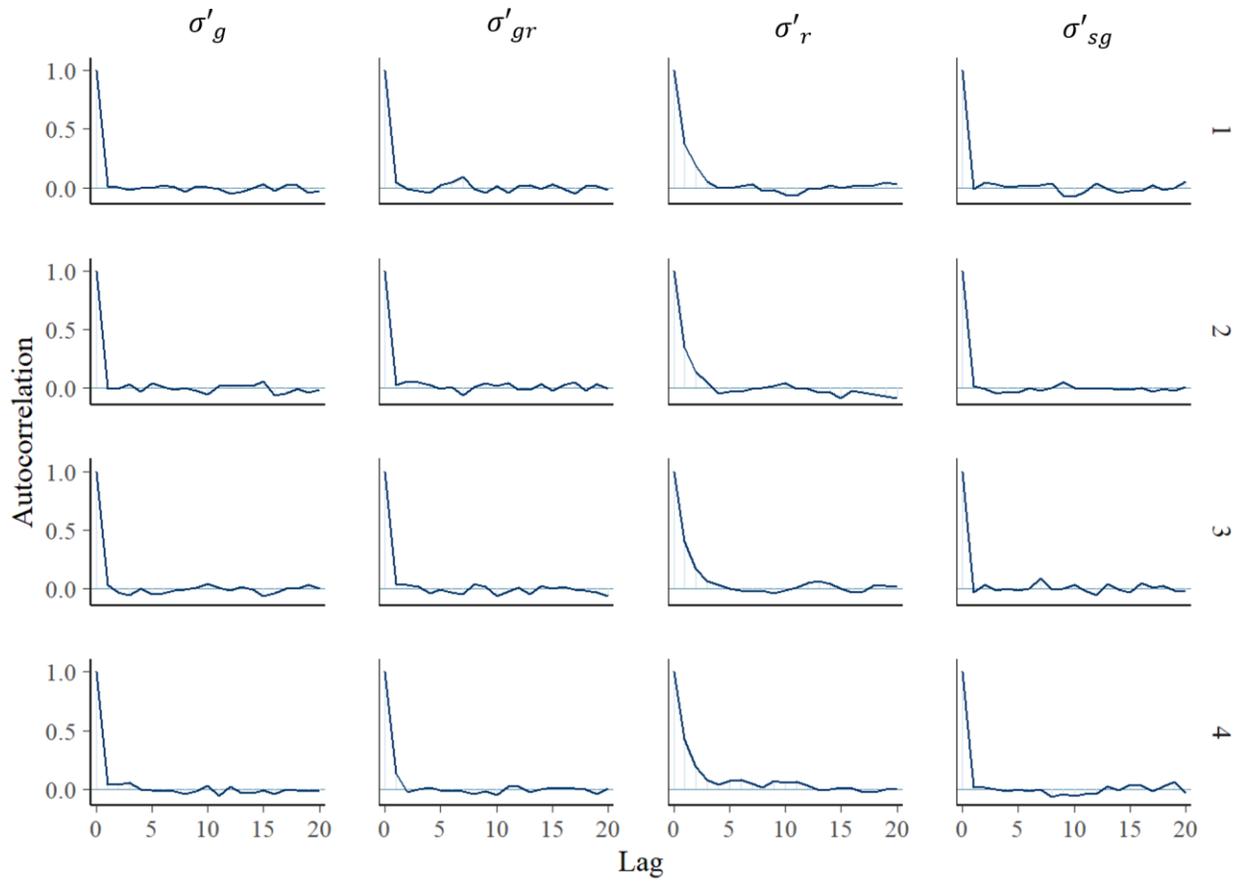

**5. Trace plots of variance components when the random dyad effect is included in the model**

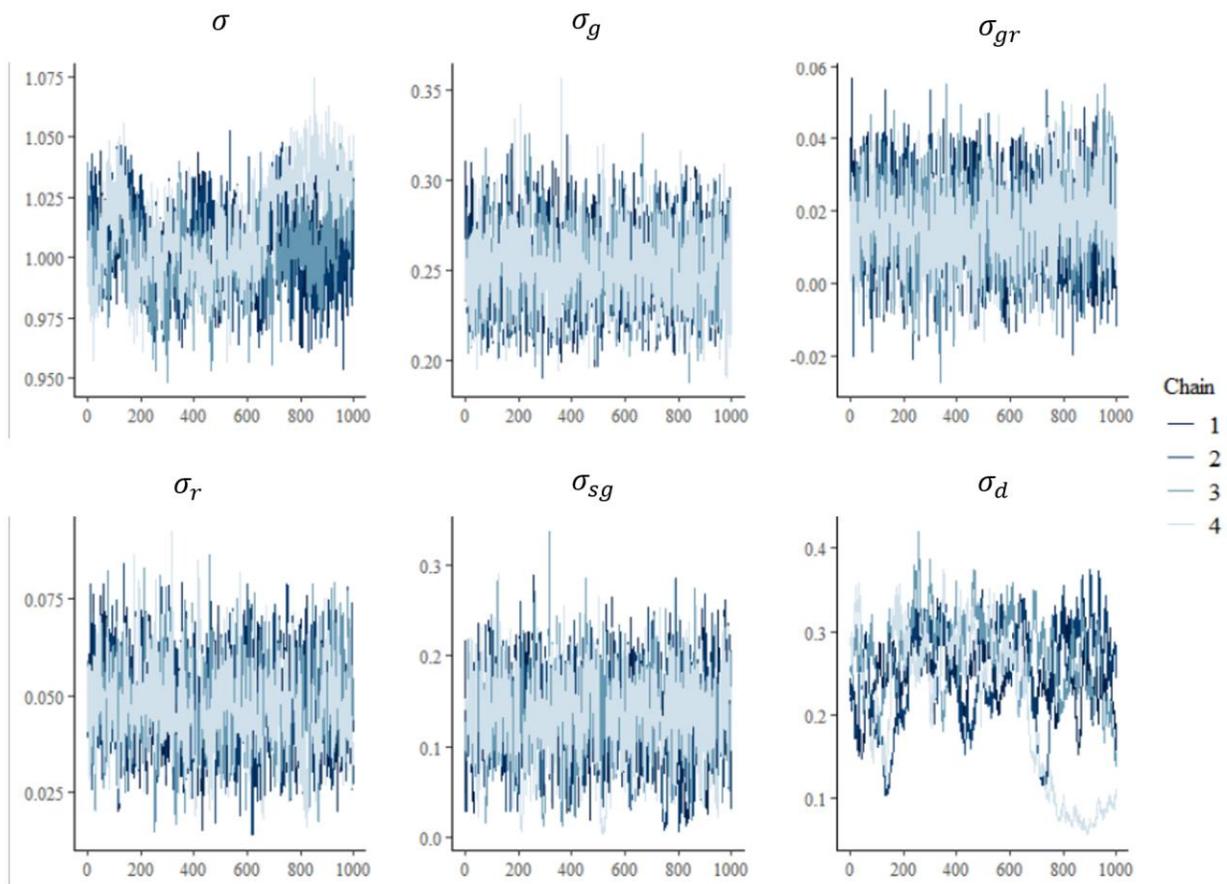

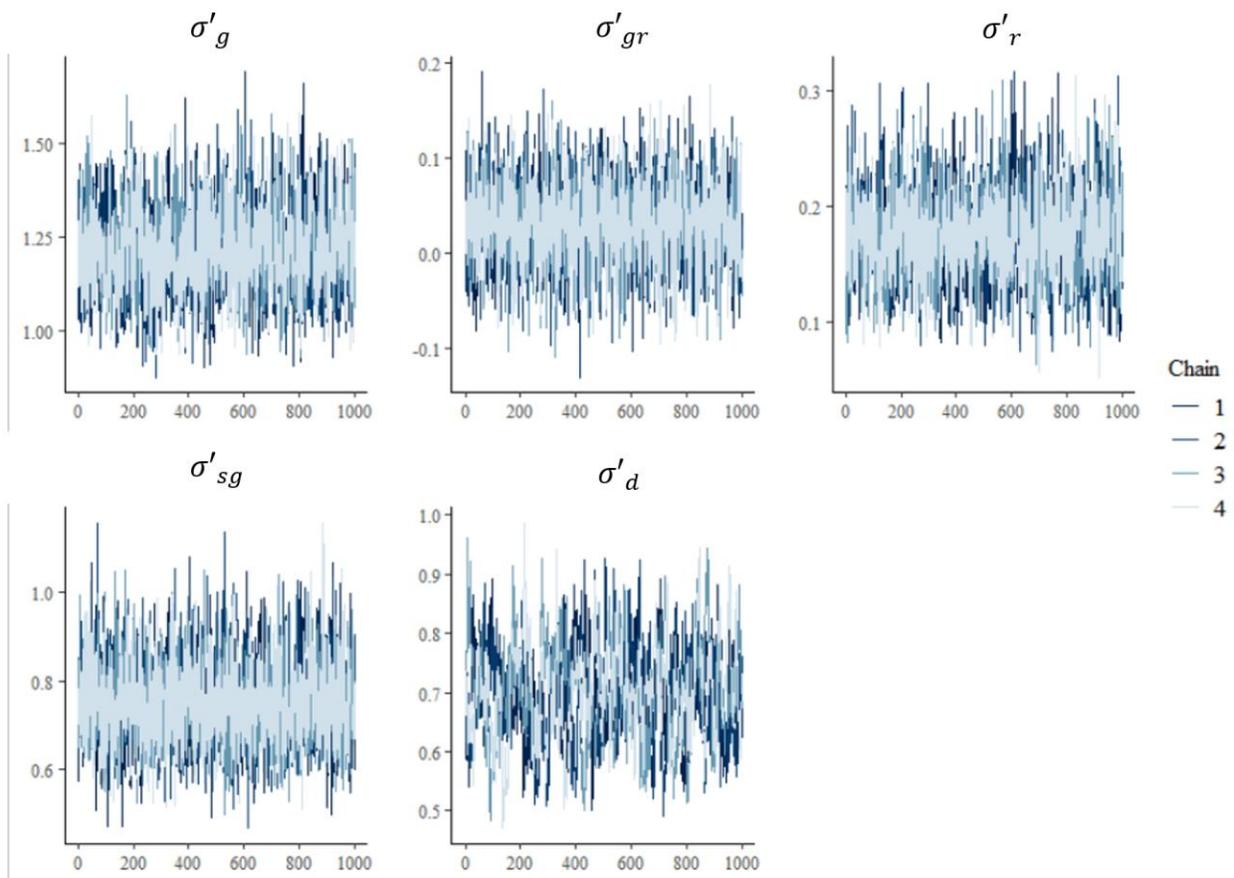

## 6. Autocorrelation plots by chain and by variance components when the random dyad effect is included in the model

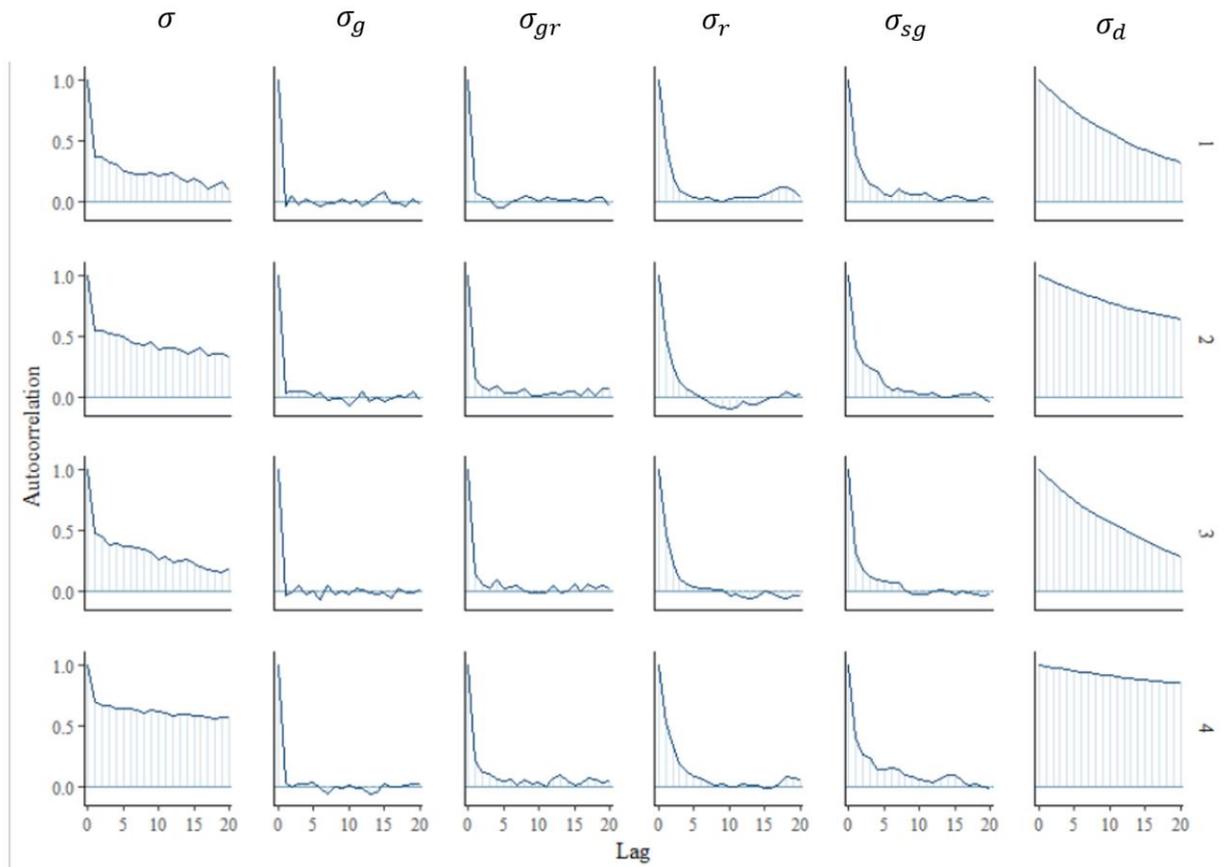

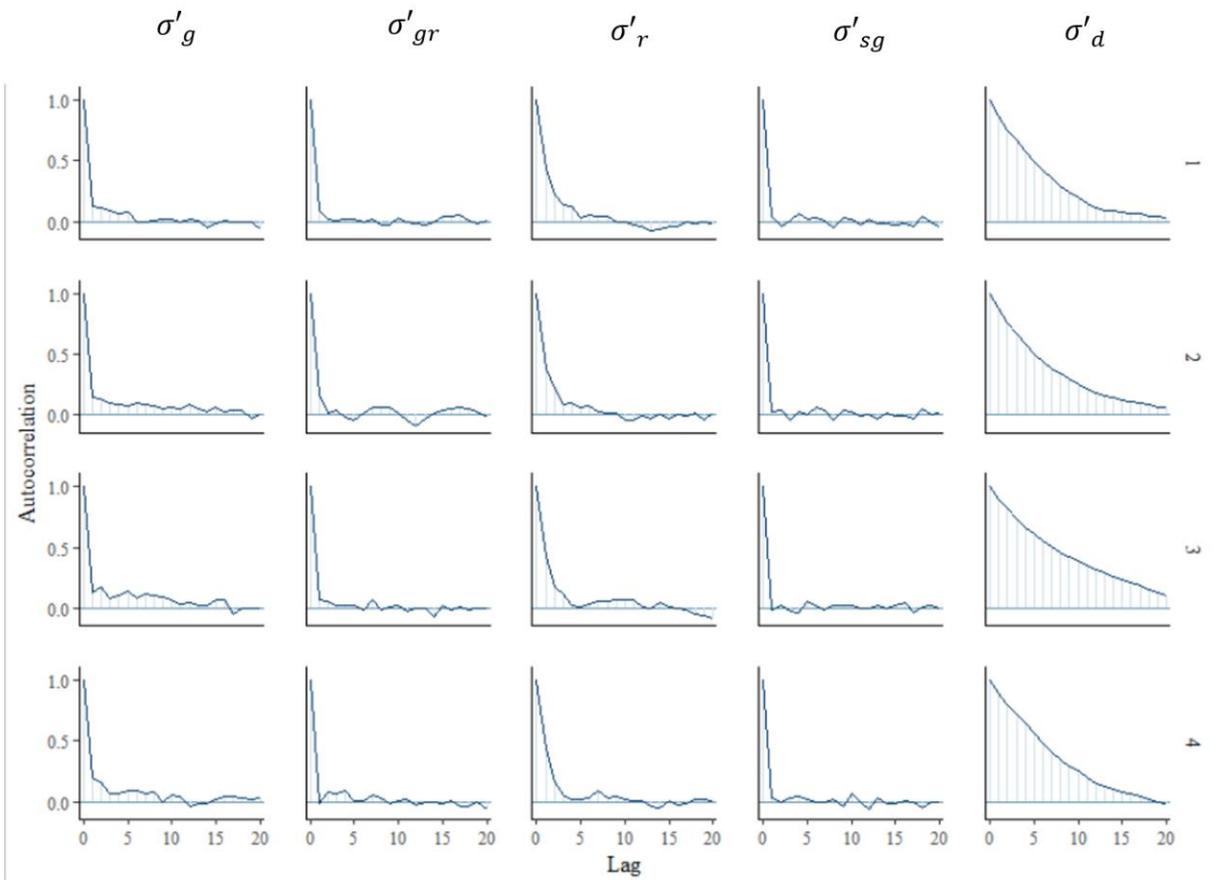

**7. Summary of MCMC samples. Q, quantile; n_eff, effective sample size.**

| Parameter | mean | sd | Q2.5% | Q50% | Q97.5% | n_eff | Rhat |
|---|---|---|---|---|---|---|---|
| $sex$ | -0.038 | 0.063 | -0.159 | -0.038 | 0.088 | 3711.37 | 1.002 |
| $\delta_1$ | -0.502 | 0.038 | -0.574 | -0.501 | -0.427 | 4127.503 | 1.000 |
| $\delta_2$ | -0.023 | 0.059 | -0.14 | -0.023 | 0.095 | 3887.644 | 1.000 |
| $\alpha$ | -0.001 | 0.004 | -0.009 | -0.001 | 0.006 | 3316.333 | 1.000 |
| $\beta$ | -0.007 | 0.005 | -0.018 | -0.007 | 0.004 | 3820.173 | 1.000 |
| $\sigma$ | 1.006 | 0.018 | 0.973 | 1.005 | 1.042 | 35.164 | 1.102 |
| $\sigma_{sg}$ | 0.134 | 0.048 | 0.030 | 0.135 | 0.223 | 954.933 | 1.000 |
| $\sigma_d$ | 0.245 | 0.066 | 0.076 | 0.258 | 0.340 | 17.190 | 1.232 |
| $sex'$ | -0.096 | 0.218 | -0.521 | -0.094 | 0.337 | 3921.573 | 1.000 |
| $\delta'_1$ | 0.552 | 0.068 | 0.420 | 0.552 | 0.684 | 3572.37 | 1.000 |
| $\delta'_2$ | -0.189 | 0.109 | -0.402 | -0.19 | 0.027 | 3438.372 | 1.001 |
| $\alpha'$ | 0.001 | 0.007 | -0.012 | 0.001 | 0.014 | 4097.341 | 1.000 |
| $\beta'$ | -0.009 | 0.011 | -0.029 | -0.009 | 0.012 | 3935.423 | 1.002 |
| $\sigma'_{sg}$ | 0.749 | 0.094 | 0.584 | 0.744 | 0.954 | 3759.123 | 1.000 |
| $\sigma'_d$ | 0.697 | 0.079 | 0.548 | 0.697 | 0.855 | 266.248 | 1.004 |
| $\sigma_r$ | 0.047 | 0.011 | 0.026 | 0.047 | 0.071 | 1281.134 | 1.001 |
| $\sigma'_r$ | 0.173 | 0.039 | 0.103 | 0.172 | 0.253 | 1341.962 | 0.999 |
| $\sigma_g$ | 0.251 | 0.022 | 0.212 | 0.250 | 0.295 | 3503.677 | 0.999 |
| $\sigma'_g$ | 1.216 | 0.113 | 1.011 | 1.209 | 1.450 | 1264.305 | 1.001 |
| $\rho_{gr}$ | 0.147 | 0.107 | -0.066 | 0.145 | 0.360 | 1322.434 | 1.008 |
| $\rho'_{gr}$ | 0.068 | 0.093 | -0.12 | 0.071 | 0.244 | 2482.093 | 0.999 |
| lp__ | -17317.6 | 1792.383 | -19481.7 | -17766.3 | -11893.2 | 15.225 | 1.270 |